# Random Sampling Using Shannon Interpolation and Poisson Summation Formulae

Xiao Z. Wang[a] and Wei E.I. Sha[b]

a. Department of information Science & Electronic Engineering, Zhejiang University, No. 38 Zheda Road, Hangzhou, Zhejiang Province, China. Email: wxzwwm@yahoo.com

b. Department of Electrical and Electronic Engineering, the University of Hong Kong, Pokfulam Road, Hong Kong. Email: wsha@eee.hku.hk

**Abstract:** This report mainly focused on the basic concepts and the recovery methods for the random sampling. The recovery methods involve the orthogonal matching pursuit algorithm and the gradient-based total variation strategy. In particular, a fast and efficient observation matrix filling technique was implemented by the classic Shannon interpolation and Poisson summation formulae. The numerical results for the trigonometric signal, the Gaussian-modulated sinusoidal pulse, and the square wave were demonstrated and discussed. The work may give some help for future work in theoretical study and practical implementation of the random sampling.

## I. Random Sampling versus Compressive Sensing

### A. *Short introduction of compressive sensing*

The well-known Shannon sampling theorem that the sampling frequency must be at least twice the maximum frequency of the signal has dominated the information theory for many years. Recently, the compressive sensing theory has broken the rule and drawn a lot of attentions from scientists and engineers. The basic idea underlying the theory is that the sparse signals can be reconstructed from generally incomplete non-adaptive information. The two fundamental features for the compressive sensing are sparsity and incoherence. The sparsity gives the opportunity that data can be under-sampled, while incoherence results in the nearly perfect reconstruction from the random measurements. (For more details, one can refer to [1] and [2].)

We consider a discrete signal $x_d^u$ of length $N$ (Normally, the signal is uniformly sampled), and a random observation matrix $\mathbf{M_0}$ of size $M \times N$. The encoding procedure is given by

$$y_d = \mathbf{M_0} x_d^u \qquad (1)$$

where $y_d \in R^{M \times 1}$ is the measurement. If $x_d^u$ is sparse in the basis system $\mathbf{\Psi}$, the decoding procedure comes down to a regularized recovery method, i.e.

$$\min \| \tilde{x}_d^u \|_{l_1} \quad \text{s.t.} \quad y_d = \mathbf{M_0} \mathbf{\Psi}^* \tilde{x}_d^u \qquad (2)$$

where $\tilde{x}_d^u = \mathbf{\Psi} x_d^u$ and $\mathbf{\Psi}$ is the transform matrix (Fourier transform, wavelet transform, etc), which satisfies the relation $\mathbf{\Psi}\mathbf{\Psi}^* = \mathbf{\Psi}^*\mathbf{\Psi} = \mathbf{I}$. The $l_1$ norm regularization problem can be solved by a variety of recovery methods involving the orthogonal matching pursuit (OMP) algorithm [3] and the gradient-based total-variation (TV) strategy [4].

B.  *Short introduction of random sampling*

Different from the compressive sensing, the measurement of the random sampling is obtained by the non-uniform sampling. The unique feature of the random sampling paves the way for its hardware implementation. According to the theory of the random sampling, the continuous analog signal can be converted to the discrete digital signal by an analog-to-digital converter with the random sampling interval [5].

Traditionally, if the uniform sampling frequency is two times larger than the maximal frequency of the original continuous band-limited signal $x_c(t)$, the continuous signal can be perfectly reconstructed from its samples $x_d^u[n] = x_c(nT)$.

However, the random sampling can achieve lower "average" sampling frequency for perfectly recovering the continuous signal if it is sparse in one domain. To some extent, the random sampling breaks the Shannon sampling theorem by using the non-uniform sampling. In many realistic situations, the random sampling becomes very powerful due to the data-missing problems, sometimes due to practical limits. Taking the medical imaging for example, the computerized tomography (CT) and the magnetic resonance imaging (MRI) frequently use the random polar and spiral sampling sets [6]. In physics and chemistry, fewer measurements can be taken at random to reduce the repeat of the experiments. That's the reason why the random sampling can make a difference to the science and engineering.

To begin with, Candès, Romberg and Tao proved that by the random sampling in the Fourier domain, one can recover the data exactly with high probability [1]. Then the conclusion has been generalized by Kunis and Rauhut [7] [8] for the samples taken at random from the cube $[0, 2\pi]^d$. After this, the random sampling was applied to solve some engineering problems.

## II. Theory of Random Sampling

Given a time-domain continuous signal $x_c(t)$, the discrete random measurement $x_d^r \in R^{M \times 1}$ is taken at the uniformly distributed time sequence $T = \{t_1, t_2, \cdots t_M\}$, i.e.

$$x_d^r[m] = x_c(t_m), \quad m = 1, 2, \cdots M, \quad t_1 < t_2 < \cdots t_M \tag{3}$$

We aim at recovering the uniformly "over-sampled" data $x_d^u \in R^{N \times 1}$ ($N >> M$) from the random "under-sampled" measurement $x_d^r$. If $x_d^u$ is sparse in the transform domain $\Psi$, then the recovery method is given by

$$\min \| \tilde{x}_d^u \|_{l_1} \quad \text{s.t.} \quad x_d^r = \mathbf{M_0} \Psi^* \tilde{x}_d^u \tag{4}$$

The above is very similar to the recovery method of the compressive sensing (2). However, the observation matrix $\mathbf{M_0}$, which connects the uniformly sampled data $x_d^u$ with the random measurement $x_d^r$, should be constructed by the well-known Whittaker-Shannon interpolation formula

$$\mathbf{M_0}(m, n) = \sin c\left(\frac{t_m}{T} - n\right), \quad 1 \leq m \leq M, \quad 1 \leq n \leq N \tag{5}$$

where $T$ is the uniform sampling interval of $x_d^u$.

Unfortunately, using the OMP recovery algorithm, we cannot obtain a good recovered result. Because the discretization in the frequency domain leads to the periodic extension in the time domain and vice versa. When we use the discrete Fourier transform, the finite-length sequence is regarded as the periodic sequence. In other words, the implicit periodic property is inherent in the discrete Fourier transform. In (5), only one period contribution is considered while the contributions from other periods are ignored. The function $\sin c(x)$ damps as the order of $\frac{1}{x}$, thus (5) may play the most important role. However, $\sin c(x)$ function is not compact support. As a result, the contributions from other periods cannot be ignored.

Based on the above analyses, (5) can be revised as

$$\mathbf{M}_0(m,n) = \sum_{p=-\infty}^{+\infty} \sin c\left(\frac{t_m}{T} - n + pN\right) \approx \sum_{p=-\frac{P}{2}+1}^{\frac{P}{2}} \sin c\left(\frac{t_m}{T} - n + pN\right) \quad (6)$$

Numerically, we truncate the summation with P terms. However, the summation itself converges vey slowly and a large number of terms should be used for obtaining accurate result. To accelerate the summation, the Poisson summation formula is employed, i.e.

$$\sum_{p=-\infty}^{\infty} f_c(t+pN) = \frac{1}{N}\sum_{k=-\infty}^{\infty} \tilde{f}_c\left(\frac{k}{N}\right)\exp\left(2\pi j\frac{k}{N}t\right) \quad (7)$$

where $\tilde{f}_c$ is the continuous Fourier transform of $f_c$. We know that the continuous Fourier transform of $\sin c(x)$ is the rectangular pulse with finite support. From (7), (6) can be rewritten as

$$\mathbf{M}_0(m,n) = \frac{1}{N}\sum_{k=-\frac{N}{2}+1}^{\frac{N}{2}} \exp\left[2\pi j\frac{k}{N}\left(\frac{t_m}{T} - n\right)\right] \quad (8)$$

The above uses the finite-term summation and is a geometric progression, which can be evaluated very fast. With the help of the Poisson summation formula, we transform the infinite summation in the time domain to the finite summation in the frequency domain. According to our numerical results below, the observation matrix constructed by (8) achieves very accurate and efficient numerical performances.

### III. Numerical Results

*A. Trigonometric signal*

We take the following signal

$$x_c(t) = 0.3\sin(2\pi f_1 t) + 0.6\cos(2\pi f_2 t) + 0.1\sin(2\pi f_3 t) + 0.9\cos(2\pi f_4 t) \qquad (9)$$

as an example, where $f_1$=50Hz, $f_2$=100Hz, $f_3$=200Hz, and $f_4$=400Hz. The number of the random samples is set to 64 and the OMP algorithm is employed. The length of the recovered signal is set to N=256 and the critical sampling rate of $f_s$=800Hz is adopted.

First, we only consider one period contribution by using (5). The program is run for 50 times, and the average reconstruction error is calculated. Figure 1 shows the recovered signal of one reconstruction. We can see that the recovered result is poor and the average reconstruction error (relative two-norm error) is 37.81%.

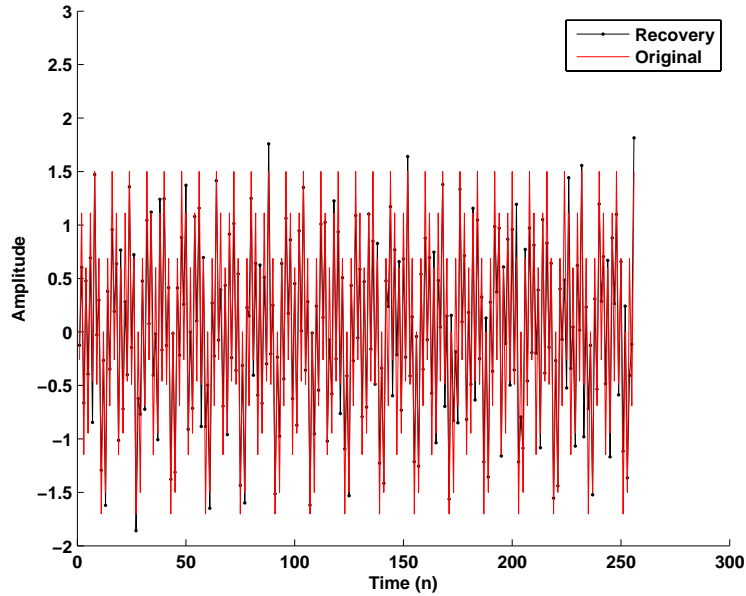

Fig. 1. The recovered trigonometric signal by the random sampling and the OMP algorithm. The observation matrix is constructed by using (5).

Second, we consider the contributions from 200 periods by using (6). Likewise, the program is run for 50 times. Figure 2 shows the result of one reconstruction and we can see that the recovered result is good. Even so, the average reconstruction error is still about 1%, which is not small enough.

Third, as shown in Fig. 3 and Fig. 4, the average reconstruction error and CPU time are drawn versus the number of the truncation terms P in (6). From Fig.3, the reconstruction error converges very slowly by using the time-domain summation equation (6). On the contrary, the error by the Poisson summation formula (8) is $2.14 \times 10^{-14}$. As shown in Fig. 4, when the number of the truncation terms P increases, the CPU time increases also. For comparison, the average CPU time by the Poisson summation formula is 0.1258 seconds.

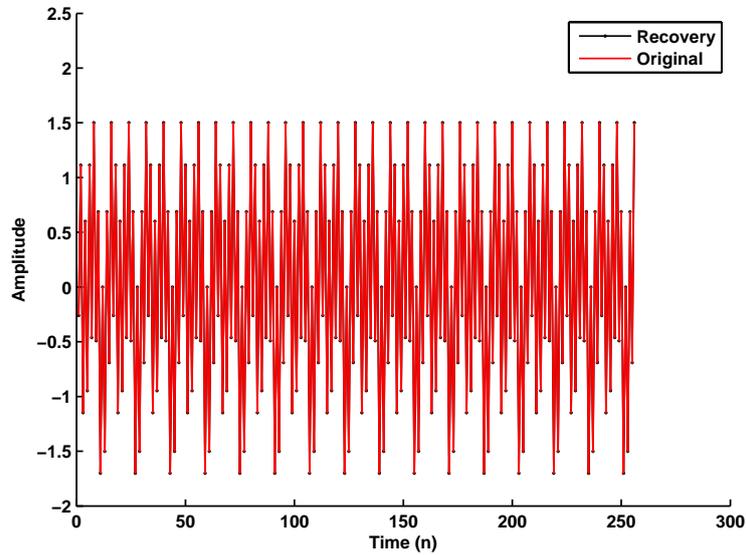

Fig. 2. The recovered trigonometric signal by the random sampling and the OMP algorithm. The observation matrix is constructed by using (6) and P=200.

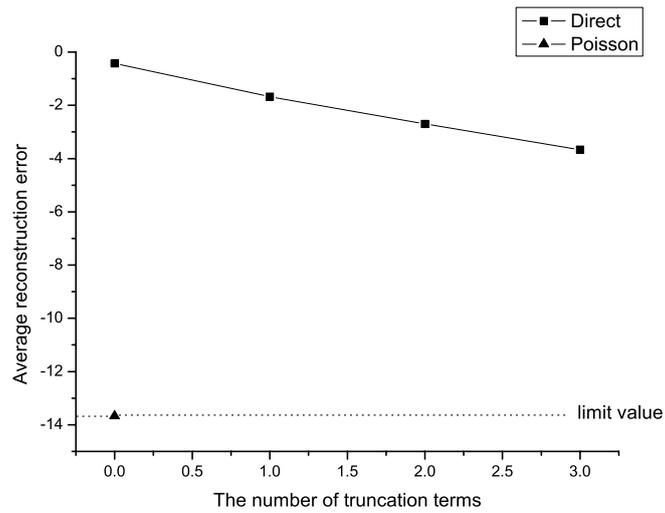

Fig. 3. The average reconstruction error (solid line) versus the number of the truncation terms. The dotted line denotes the error by the Poisson summation formula (8). The logarithmic scales are used for both x and y coordinates.

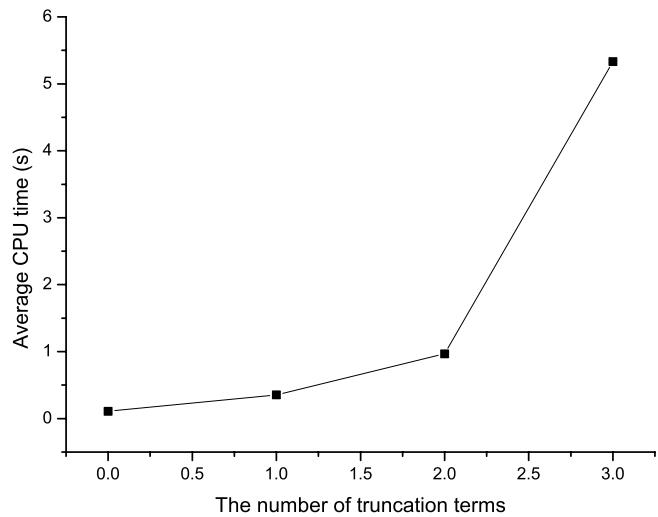

Fig. 4. The average CPU time versus the number of the truncation terms. The logarithmic scale is used for x coordinate only.

*B. Gaussian-modulated sinusoidal pulse*

The recovered signal is a 50 kHz Gaussian-modulated sinusoidal pulse with 60% bandwidth which is sampled at a rate of 10 MHz and truncated where the envelope falls 60 dB below the peak. M is set to 93 and N is set to 928. The observation matrix is generated by the Poisson summation formula (8). The OMP algorithm is employed again. Figure 5 shows the result of one reconstruction. The average reconstruction error is 1.44% and the average CPU time is 2.0366 seconds. The reconstruction is not perfect, because we only recover the most important components in the Fourier domain and set others to zero.

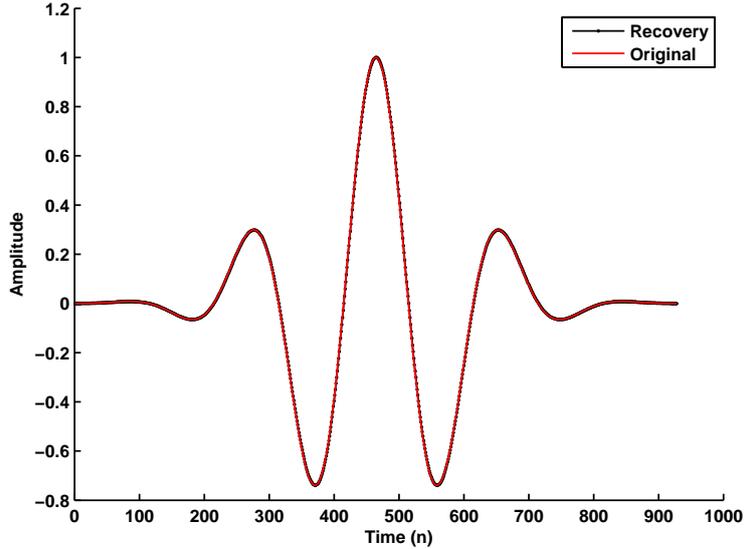

Fig. 5.  The recovered Gaussian-modulated sinusoidal pulse by the random sampling and the OMP algorithm. The observation matrix is constructed by using (8).

*C. Square wave*

The square wave is not sparse in the Fourier domain, but its variation in the time domain is sparse. In this case, we employ the gradient-based TV strategy as the recovery method. The square wave is extended to a periodic signal and the Poisson summation formula (8) is used. However, the periodic extension will introduce an overshoot if the borders of the signal have different amplitudes. Figure 6 shows the result of one reconstruction. M is set to 80 and N is set to 240. The recovered result breaks down at the sharp edges mainly due to the fact that the square wave contains infinite time harmonics and the Gibbs phenomenon appears. The Shannon interpolation formula cannot overcome the Gibbs phenomenon and the observation matrix $\mathbf{M}_0$ produced is not accurate.

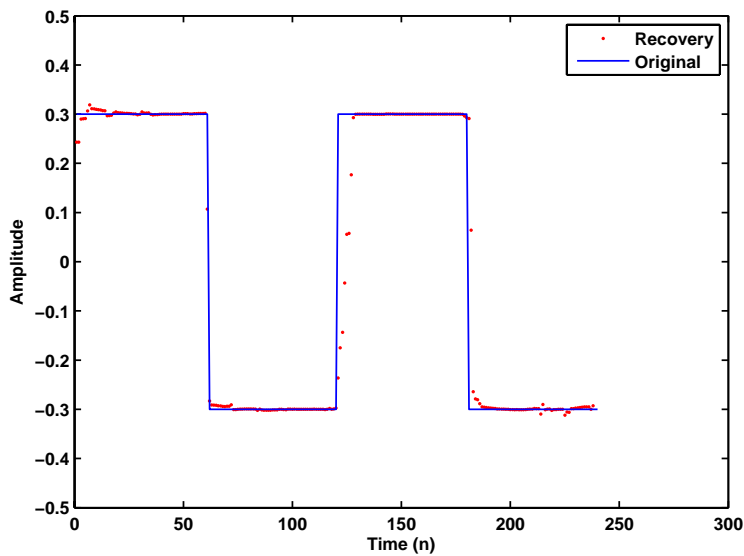

Fig. 6. The recovered square wave by the random sampling and the gradient-based TV strategy. The observation matrix is constructed by using (8).

## IV. Conclusion

In this report, we discussed the basic theory of the random sampling and its recovery methods. Different from the compressive sensing, the observation matrix of the random sampling is constructed by the Shannon interpolation formula. Due to the implicit periodic property of the discrete Fourier transform, the infinite summation is required for the interpolation formula. By the aid of the Poisson summation formula, the infinite summation in the time domain successfully converts to the finite summation in the frequency domain, which converges much faster. The numerical results demonstrated the accuracy and efficiency of the proposed methods. The future work can focus on other interpolation formulae or methods, such as wavelet interpolation algorithm.